\documentstyle[epsfig]{aipproc}

\begin{document}
\title{Stretched Horizon for Non--Supersymmetric Black Holes }

\author{C. Espinoza and M. Ruiz--Altaba}
\address{Departamento de F\'{\i}sica Te\'orica\\
Instituto de F\'{\i}sica \\
Universidad Nacional Aut\'onoma de M\'exico\\
Apartado Postal 20-364\\
01000 M\'exico, D.F.}

\maketitle

\begin{abstract}     
We review the idea of stretched horizon for extremal black holes in supersymmetric string theories, and we compute it for non-supersymmetric black holes in four dimensions. Only for small masses of the order of the Veneziano wavelength is the stretched horizon bigger than the event horizon.
\end{abstract}

\section{Introduction}

It was long ago suggested that black holes should be treated as elementary particles \cite{tHooft}, because both are parametrized only by their mass, angular momentum (or spin) and gauge charges. String theory has made a significant contribution towards putting  this assertion on a firm basis. A  common feature of black holes and elementary string states is that   the degeneracy of states with a given mass increases with it. Unfortunately, for elementary string states the logarithm of the degeneracy of states increases linearly with mass, whereas the Bekenstein-Hawking entropy of the black hole increases as the square of the mass. There are some cases in which the discrepancy between the two entropies can be removed appealing to the large renormalization of the mass of a black hole\cite{Sus1}. 
There are, however, some particular states in string theory, called BPS states, which do not receive any mass renormalization\cite{Witten}.   Whereas the logarithm of the degeneracy of BPS states grows linearly with the mass, the area of the event horizon of a BPS or  extremal black hole actually vanishes. This motivates the argument\cite{sen} that the entropy of an extremal black hole is not exactly equal to the area of the event horizon, but to the area of a surface close to the event horizon called ``stretched horizon''. By carefully defining the location of the stretched horizon in a consistent way, the Bekenstein-Hawking formula for the black hole entropy can receive corrections in such a way that it correctly reproduces the logarithm of the density of elementary string states.

The stretched horizon of a black hole is defined as the surface where the space-time curvature (in the string metric) becomes large.  
It is also the surface where the local Unruh temperature for a stationary observer (constant $r$)  is  the Hagedorn temperature of the string theory. (The local Unruh temperature becomes infinite at the event horizon).  
Specifically, we define the stretched horizon as the surface where the scalar curvature $C=($Riemann$)^2$ is equal to one in Planck's units.

\section{Stretched horizon for the classical solutions}

Throughout, we shall be thinking of black holes as the classical description of a quantum object. The nature of this object is well approximated by (classical) general relativity far away from the horizon. At any rate, we should not expect the metric which solves Einstein's equations to make any sense at distances to the origin smaller than the de Broglie wavelength $\lambda_B= M^{-1}$ for a black hole of mass $M$.
Since we are stringy, it is perhaps more appropriate to use Veneziano's generalized uncertainty principle\cite{Veneziano}  $\Delta x \geq \frac{\ell_s^2}{2\hbar} \Delta p + \frac{\hbar}{ \Delta p}$, where $ \ell_s$ is the string scale,   $g$ is the string coupling constant and $\ell_p= g \ell_s$ is Planck's length. 
So, in Planck units, the metric certainly is not expected to make any sense at radii smaller than 
\begin{equation}{\label {veneciano} }
\lambda_V = {\frac1 2} M + {\frac1 M}
\end{equation}

Where is the stretched horizon? If the place where the curvature becomes big (unity in Planck units) is inside the event horizon, then clearly there is no need to stretch it at all. Similarly, if it falls at a radius smaller than $\lambda_V$, there is no point in talking about it.  We are interested in finding under what circumstances the stretched horizon is a meaningful and useful concept for non-supersymmetric four-dimensional black holes. In other words, we must find when the stretched horizon is bigger than the event horizon and also bigger than $\lambda_V$ (or $\lambda_B$).

Consider the general solution to the Einstein-Maxwell equations
\begin{eqnarray}\label{metric}
ds^2 = -\left (\frac { {\Delta} - a^2{\sin^2{\theta}}} {\Sigma}\right) dt^2 - \frac{2a \sin^2{\theta}\left( r^2 + a^2 - {\Delta} \right)}{\Sigma} dtd{\phi}\,\,\,\,\,\,\, \nonumber \\
+ \left[ \frac {{\left( r^2 + a^2 \right)}^2 - {\Delta}a^2 \sin^2{\theta}}{\Sigma}\right] \sin^2 {\theta} d{\phi}^2 + \frac {\Sigma}{\Delta} dr^2 + {\Sigma}{d{\theta}}^2,
\end{eqnarray}
\begin{equation}
A_{\mu} = - \frac {er}{\Sigma} \left[ {(dt)}_{\mu} - a\sin^2 {\theta}({d{\phi}}_{\mu}) \right],
\end{equation}
 where
\begin{equation}
{\Sigma} = r^2 + a^2cos^2{\theta}, \,\,\,\,\,\, {\Delta} = r^2 + a^2 + e^2 - 2Mr,
\end{equation}
Using Mathematica, we have computed the square of the Riemann tensor for this solution, $C$, and evaluated where it becomes one. 

For the Schwarszchild black hole ($e=a=0 $) the scalar curvature is simply
\begin{equation}{\label {sch} }
C = \frac{48 M^2}{r^6}
\end{equation}
whereby the stretched horizon radius is 
\begin{equation}{\label {rsch} }
r_s = 48^{\frac{1}{6}} M^{\frac{1}{3}}
\end{equation}
Thus, the stretched horizon is always inside the event horizon, except for such ridiculously small masses that the de Broglie wavelength is actually bigger than both.  

This continues to be the case for any charged static black hole: in Fig.~1 we plot the critical mass (in Planck units) at which the stretched horizon crosses the event horizon in terms of the  charge $e$. Only for masses below the line is the stretched horizon bigger than the event horizon. The curve is essentially flat just below $M\sim 0.95$  when it begins to grow. When $e$ reaches its extremal value $e=1$, $M_c$ reaches 1.09.

So both for Schwarzschild and for Reissner-Nordstr\"om black holes, the stretched horizon is useless.

When $a\ne0$, i.e. for rotating black holes, the situation doesn't change much. The stretched horizon is again bigger than the event horizon only for small masses, but the critical mass below which the stretched horizon is relevant (bigger than the event horizon) is now a few times bigger and grows with $a$. Fig.~2 shows this critical mass when $e=0$ (the plot was computed in the axial direction $\theta=\pi$). In the extremal limit $a\to1$, the critical mass approaches the value $5.5$ corresponding to a stretched horizon radius of $11.0$, which is $3.7$ times bigger than Veneziano's wavelength and $63.9$ times bigger than de Broglie's wavelength.

 The charge, in the general Kerr-Newman case, remains rather irrelevant.

\section{Conclusions}
In this work we have shown two things. First, that outside the event horizon of a static black hole, the curvature is always small. Secondly, we have found numerically  the critical mass above which the event horizon is bigger than the stretched horizon for a four dimensional black hole. In the extremal limit, it ranges from  1.09 when $a=0$, to $5.5$ when $e=0$. This results are in general consistent, since the temperature near the horizon of a classical black hole is always much smaller than the corresponding temperature of an extremal supersymmetric black hole. Since the critical mass is of order unity in all cases this mean that the stretched horizon is not a very useful concept for four-dimensional non-supersymmetric black holes. Physically, this indicates that only the massless modes of the string can be seen by an outside observer.

\begin{figure}[t!] 
\centerline{\epsfig{file=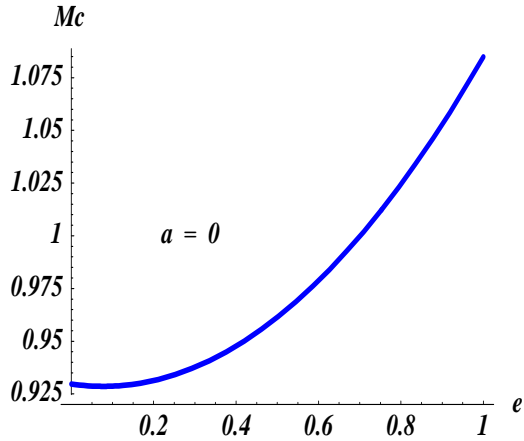,height=2.0in,width=2.5in}}
\vspace{10pt}
\caption{Critical mass $Mc$ for a nonrotating charged black hole.}
\label{fig1}
\end{figure}

\bigskip
\bigskip
\bigskip
\bigskip

\qquad

\begin{figure}[t!] 
\centerline{\epsfig{file=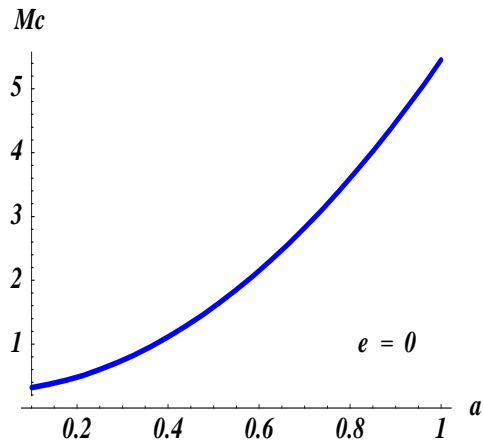,height=2.0in,width=2.5in}}
\vspace{10pt}
\caption{Critical mass $Mc$ for a rotating black hole.}
\label{fig2}
\end{figure}

\bigskip
\bigskip
\bigskip
\bigskip

{\bf Acknowledgements}. This work is supported in part by CONACYT 25504E,  DGAPA-UNAM IN103997. C.E. enjoys a scholarship from DGEP-UNAM.

\end{document}